\documentclass[pra,aps,twocolumn,nopacs,superscriptaddress,nofootinbib,longbibliography]{revtex4-1} 
\usepackage{amsmath}  \usepackage{amssymb}  \usepackage{amsfonts}  \usepackage{bm}  \usepackage{bbm}   \usepackage{braket}  \usepackage{color}  \usepackage{comment}  \usepackage{dcolumn}  \usepackage{enumerate}  \usepackage{epsfig}  \usepackage{gensymb}  \usepackage{graphicx}  \usepackage{indentfirst}  \usepackage{lmodern}  \usepackage{mathrsfs}  \usepackage{mathtools}  \usepackage{psfrag}  \usepackage{pst-all}  \usepackage{soul}  \usepackage{units}  \usepackage{xcolor}
\usepackage{float} 
\usepackage[colorlinks,linkcolor=blue,citecolor=blue,urlcolor=blue,hyperindex,driverfallback=dvipdfm]{hyperref}  \usepackage[T1]{fontenc} 

\def\ii{{\rm i}}  \def\ee{{\rm e}}
\def\me{m_{\rm e}}  \def\kB{{k_{\rm B}}}
\def\Ree{{\rm Re}}  \def\Imm{{\rm Im}}
                              \def\kb{{\bf k}}              \def\Rb{{\bf R}}  \def\rb{{\bf r}}      \def\vb{{\bf v}} 
                
  \def\kparb{{\bf k}_\parallel} 
\def\EF{{E_{\rm F}}}     
      
\def\Te{T_{\rm e}}  \def\Tl{T_{\ell}}
\def\ce{c_{\rm e}}

\begin{document} 
\def\bibsection{\section*{\refname}} 

\title{Generation and control of localized terahertz fields in photoemitted electron plasmas}

\author{Eduardo~J.~C.~Dias}
\email{eduardo.dias@icfo.eu}
\affiliation{ICFO-Institut de Ciencies Fotoniques, The Barcelona Institute of Science and Technology, 08860 Castelldefels (Barcelona), Spain}

\author{I.~Madan}
\author{S.~Gargiulo}
\affiliation{Institute of Physics, École Polytechnique Fédérale de Lausanne, Lausanne, 1015, Switzerland}

\author{F.~Barantani}
\affiliation{Institute of Physics, École Polytechnique Fédérale de Lausanne, Lausanne, 1015, Switzerland}
\affiliation{Department of Quantum Matter Physics, University of Geneva, 24 Quai Ernest-Ansermet, Geneva, 1211, Switzerland}

\author{M.~Yannai}
\affiliation{Technion - Israel Institute of Technology, Haifa 3200003, Israel}

\author{G.~M.~Vanacore}
\affiliation{Department of Materials Science, University of Milano-Bicocca, Via Cozzi, 55, Milano, 20126, Italy}

\author{I.~Kaminer}
\affiliation{Technion - Israel Institute of Technology, Haifa 3200003, Israel}

\author{F.~Carbone}
\affiliation{Institute of Physics, École Polytechnique Fédérale de Lausanne, Lausanne, 1015, Switzerland}

\author{F.~Javier~Garc\'{\i}a~de~Abajo}
\email{javier.garciadeabajo@nanophotonics.es}
\affiliation{ICFO-Institut de Ciencies Fotoniques, The Barcelona Institute of Science and Technology, 08860 Castelldefels (Barcelona), Spain}
\affiliation{ICREA-Instituci\'o Catalana de Recerca i Estudis Avan\c{c}ats, Passeig Llu\'{\i}s Companys 23, 08010 Barcelona, Spain}

\begin{abstract}
Dense micron-sized electron plasmas, such as those generated upon irradiation of nanostructured metallic surfaces by intense femtosecond laser pulses, constitute a rich playground to study light-matter interactions, many-body phenomena, and out-of-equilibrium charge dynamics. Besides their fundamental interest, laser-induced plasmas hold potential for the generation of localized terahertz radiation pulses. However, the underlying mechanisms ruling the formation and evolution of these plasmas is not yet well understood. Here, we develop a comprehensive microscopic theory to predictably describe the spatiotemporal dynamics of laser-pulse-induced plasmas. Through detailed analysis of electron emission, metal screening, and plasma cloud interactions, we investigate the spatial, temporal, and spectral characteristics of the so-generated terahertz fields, which can be extensively controlled through the metal morphology and the illumination conditions. We further describe the interaction with femtosecond electron beams to explain recent ultrafast electron microscopy experiments, whereby the position and temporal dependence of the observed electron acceleration permits assessing the associated terahertz field. Besides its potential application to the design of low-frequency light sources, our work contributes with fundamental insight on the generation and dynamics of micron-scale electron plasmas and their interaction with ultrafast electron pulses.
\end{abstract}

\maketitle 
\date{\today} 

\section{Introduction} 

Terahertz (THz) radiation---a part of the electromagnetic spectrum sandwiched between microwaves and infrared light---has recently attracted significant attention because of its potential application in areas such as spectroscopy \cite{JCK11,KTN13,UHS11}, sensing \cite{NFK06}, imaging \cite{WKI03,DOK06,NHY07}, and communication technologies \cite{NDR16}. In this context, nanophotonics constitutes a suitable arena to test and capitalize on some unique properties of THz radiation such as the ability to penetrate through optically opaque materials \cite{WMZ08} and a high sensitivity to chemical composition \cite{WKI03,YYY13}, both of which can be manipulated through material nanostructures. However, the efficient generation of THz light remains a challenge, and even more so when aiming for nanoscale sources.

The production of THz fields typically relies on nonlinear optical phenomena such as wave mixing \cite{ZRD11,JVJ14}, optical rectification \cite{ZMJ92,RJM94,FPA10}, and frequency conversion \cite{SSK11_2,FWW19}. These methods involve simple setups fed by high-frequency optical sources such as lasers, but they generally have low efficiencies and are limited by the availability of suitable nonlinear crystals.

Electron plasmas have emerged as an appealing alternative for the generation of intense THz fields \cite{HSG93,LGF03,ZTK21}. Such plasmas can be obtained by extracting electrons from metal surfaces upon intense laser-pulse irradiation via multiphoton photoemission and thermionic emission \cite{DPV20}. If the intensity of the ionizing laser is large enough, a short-lived electron-plasma plume of a few picoseconds in duration can be formed, characterized by a high density of emitted electrons that are eventually reabsorbed by the surface or escaping away from the metal. The associated charge motion gives rise to intense, transient localized THz fields, but the precise underlying mechanisms are not yet fully understood \cite{LLL19}.

Dense electron-plasma plumes undergo a complex spatiotemporal dynamics ruled by the collective interaction among many electrons in the presence of screening by the metallic structure, thereby posing an important challenge for a comprehensive theoretical description. Nevertheless, besides their potential for application in THz technologies, the study of this phenomenon bears interest as a source of fundamental insight into the ultrafast dynamics of complex nanoscale systems, as revealed by recent experimental results obtained by employing ultrafast electron microscopes \cite{MDG23,YDG23}, whereby electron beam (e-beam) pulses are made to interact with the plasma at controlled delay times relative to the laser pulses. In fact, high-energy electrons are ideal probes for ultrafast and localized phenomena such as charged plasmas \cite{VHW18,HWV21,RB16,SSB20,CRT08} due to their ultraconfined nature, enabling a spatial/temporal resolution down to sub-nanometer/femtosecond scale, combined with a high sensitivity to electromagnetic interactions \cite{paper371}.

In this work, we study the formation and evolution of electron plasma produced upon irradiation of metal nanostructures by intense laser pulses through a parameter-free theoretical formalism that incorporates a quantitative description of electron emission, metal screening, and cloud dynamics, including electron reabsorption and the generation of localized THz fields. The process also involves substantial heating of conduction electrons in the metal, triggering ultrafast thermal dynamics that needs to be accounted for to formulate accurate predictions on the behavior of the plasma. The present model has been successfully used to explain recent experimental results on the ultrafast nanoscale spatiotemporal dynamics of electron plasmas probed by femtosecond electron pulses \cite{MDG23}. We provide a comprehensive description of the formalism and apply it to study the so-generated transient THz fields, whose duration, spatial distribution, and spectral composition are strongly dependent on the metal morphology and illumination conditions. The latter provide suitable knobs to control the THz field characteristics for potential applications.

\begin{figure*}
\centering
\includegraphics[width=0.7\textwidth]{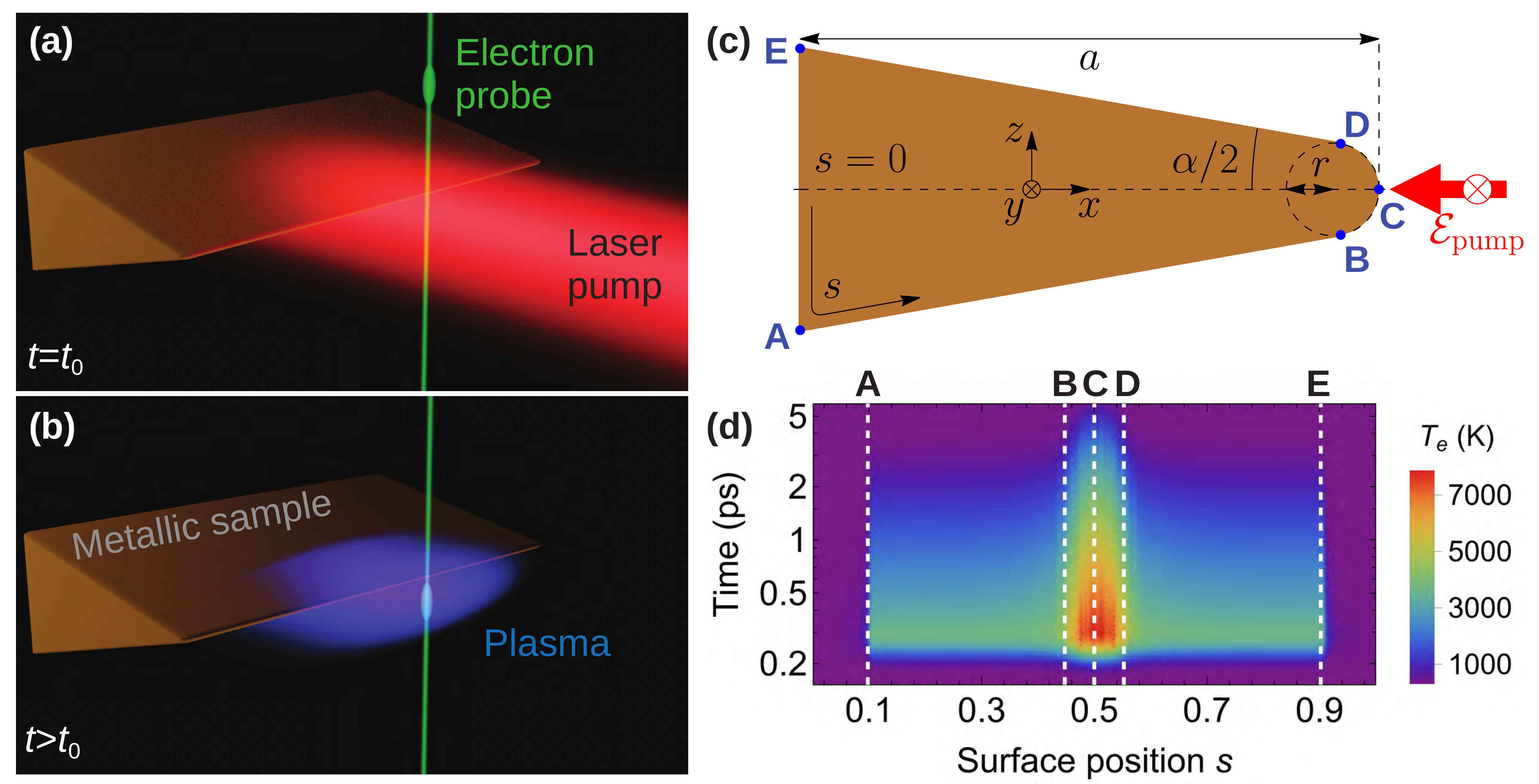}
\caption{{\bf Electron plasma formation by laser-pulse irradiation.} {\bf(a,b)} Schematic representation of (a) a metallic sample irradiated by a high-fluence laser pump pulse at time $t=t_0$ and (b) the ensuing formation of an electron plasma evolving at $t>t_0$. A synchronized e-beam probe pulse is used to study the spatiotemporal dynamics of the plasma by scanning the e-beam position and delay time relative to the laser pulse. {\bf (c)} Scheme of the specific system under consideration, consisting of a metallic wedge that is translationally invariant along $y$ and characterized by its length $a$, angle $\alpha$, and tip radius $r$, under illumination by an external field $\mathcal{E}_{\rm pump}$ polarized along $y$. The surface contour is parameterized by $s$, ranging counterclockwise from $0$ to $1$. {\bf (d)} Dynamics of the electron temperature $\Te$ at the surface of a copper wedge as a function of position $s$ and time. The vertical dashed lines mark the positions of the blue dots in (c) along the surface of the wedge, with the one at $s=0.5$ corresponding to the tip apex (point C). The wedge parameters in this specific calculation are $a=10~\mathrm{\mu m}$, $r=1~\mathrm{\mu m}$, and $\alpha=20^{\circ}$, while the laser pulse has a fluence $F_{\rm pump}=200~\mathrm{mJ/cm^2}$ and its maximum is arriving at time $t_0=0.25$~ps.}
\label{Fig1}
\end{figure*}

\section{Theoretical Description of Laser-Induced Electron Plasmas}

To demonstrate the ability of electron plumes to generate and control localized ${\rm THz}$ fields, as well as to reveal fundamental insights into their origin, we introduce a theoretical framework that describes the generation and evolution of laser-pulse-generated plasma. The theory here presented has general validity for any translationally invariant morphology and can be straightforwardly generalized to arbitrary shapes. For concreteness, we provide numerical simulations for a system composed of a translationally invariant wedge (along $y$) with the cross-sectional geometry shown in Fig.~\ref{Fig1}(c), characterized by a length $a$ along the $x$ direction, an angle $\alpha$, and a tip radius $r$. The wedge surface is described by a parameter $s$ that varies counterclockwise from $0$ to $1$, starting from the surface point opposite to the tip apex, as indicated in Fig.~\ref{Fig1}(c). We study this structure under uniform illumination from the side by a high-fluence laser pulse with an external field $\mathcal{E}_{\rm pump}$ polarized along $y$ (i.e., the direction of translational symmetry), as indicated by the red arrow in Fig.~\ref{Fig1}(c). In the simulations presented below, we consider a laser central wavelength $\lambda=800~\mathrm{nm}$, a laser pulse duration $\Delta t=60~\mathrm{fs}$, and a fluence $F_{\rm pump}=200~{\rm mJ/cm^2}$.

\subsection{Sample Temperature Dynamics} The absorption of laser energy by the metal structure raises the temperature of its conduction electrons, whose thermal distribution depends on the local field enhancement, and this in turn on the surface geometry and the illumination conditions. We model the evolution of the electron temperature $\Te(\rb,t)$ as a function of time $t$ and position $\rb$ on the metal surface using the two-temperature model described in Appendix~\ref{sec:TTM}.

An example of the surface electron temperature dynamics is presented in Fig.~\ref{Fig1}(d) for a copper wedge with geometrical parameters $a=10~\mathrm{\mu m}$, $r=1~\mathrm{\mu m}$, and  $\alpha=20^{\circ}$. The temperature is first experiencing a fast increase within $\sim 100~\mathrm{fs}$, with a hotspot concentrated near the tip apex ($s=0.5$), where it reaches $\Te\sim 8000~{\rm K}$. This is followed by a slow cooldown lasting for $\sim 5~{\rm ps}$, after which the conduction electrons return to ambient temperature. The temporal profile of the temperature along the A--E cuts in Fig.~\ref{Fig1}(d) are presented in Fig.~\ref{FigS1}, together with analogous results for a similar wedge with smaller tip radius $r=0.1~{\rm \mu m}$.

\begin{figure*}
\centering
\includegraphics[width=0.9\textwidth]{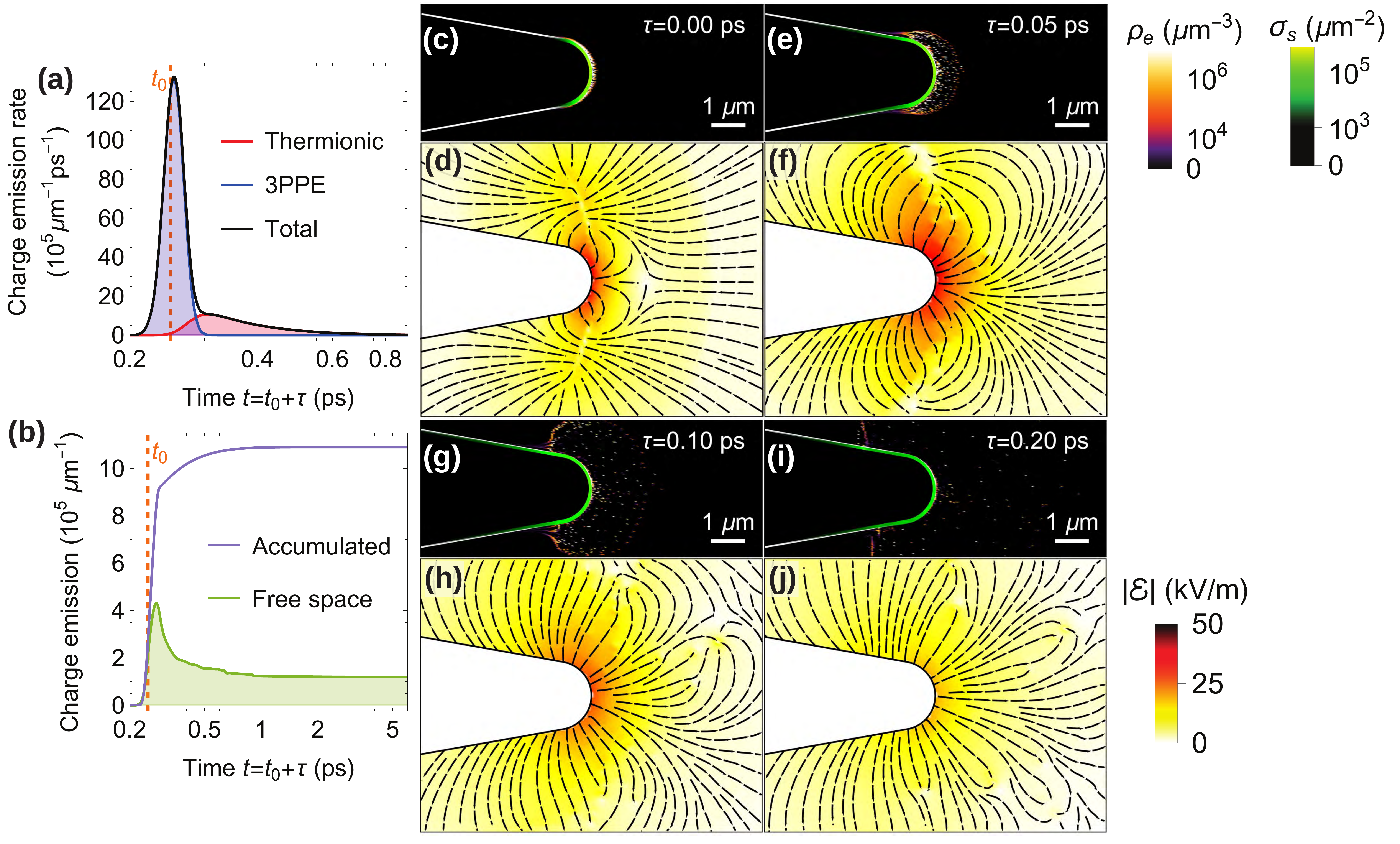}
\caption{{\bf Plasma-induced THz electric field.} {\bf (a)} Electron emission rate due to thermionic emission (red), 3PPE (blue), and both channels combined (black) as a function of time. {\bf (b)} Temporal evolution of the total number of emitted electrons (purple) and the number of electrons remaining in free space (i.e., those that are not yet reabsorbed; green) under the same conditions as in (a). {\bf (c-j)} Electron plasma density $\rho_{\rm e}$ and image-charge surface density $e\sigma_{\rm s}$ (c,e,g,i), along with the corresponding electric field maps (d,f,h,j), at four different times $t=t_0+\tau$ referred to the time $t_0$ of maximum incident pump laser intensity [indicated in panels (a,b) by an orange dashed line]. Color maps in (d,f,h,j) represent the electric-field amplitude $|\mathcal{E}|$ as a function of position, supplemented by field lines (black) tangent to $\mathbf{\mathcal{E}}$. All panels are calculated for a copper wedge with parameters $a=10~\mathrm{\mu m}$, $r=1~\mathrm{\mu m}$, and $\alpha=20^{\circ}$, as well as a pump of fluence $F_{\rm pump}=200~\mathrm{mJ/cm^2}$ arriving at $t_0=0.25~{\rm ps}$.}
\label{Fig2}
\end{figure*}

\subsection{Plasma Emission} During the time over which the metallic sample is being illuminated, as well as the subsequent period in which conduction electrons near the sample surface remain hot, substantial electron emission can occur primarily due to thermionic emission and $n$-photon photoemission (3PPE with $n=3$ in the present study, see Appendix~\ref{sec:photoemission}). The former occurs while the electron temperature remains elevated (roughly $\gtrsim1$~ps), allowing some high-energy electrons to overcome the work function $\Phi$ of the metal. In contrast, 3PPE occurs exclusively during the pumping period ($<100$~fs) via the absorption of three photons by one electron on the metal surface. These mechanisms, which are discussed in Appendix~\ref{sec:photoemission}, yield a total emission rate given by
\begin{align}
    \mathcal{P}=\mathcal{P}_{\rm th}+\mathcal{P}_{\rm 3PPE},
   \label{C6:eq:Pe:final}
\end{align}
where the electron emission rates corresponding to each mechanism, $\mathcal{P}_{\rm th}$ and $\mathcal{P}_{\rm 3PPE}$, are given by Eqs.~\eqref{eq:I:PE:Pe1:final} and \eqref{eq:I:PE:Pe2:final}, respectively. Notably, these rates are strongly dependent on the local electron temperature, which is in turn a function of position and time. 

Using Eq.~\eqref{C6:eq:Pe:final} together with the data in Fig.~\ref{Fig1}(d), we can retrieve the density of electrons $\mathcal{P}(\rb_s,t)$ emitted from a sample surface position $\rb_s$ at time $t$, which is here normalized per unit length along the direction $y$ of translational invariance in the sample. An example of total emission (i.e., integrated along the transverse surface profile) from each of the two mechanisms considered is shown in Fig.~\ref{Fig2}(a) for the copper wedge introduced in Fig.~\ref{Fig1}(d). We note that the 3PPE process is dominant during the pumping period (centered around $t_0=0.25~{\rm ps}$). In contrast, thermionic emission becomes dominant at a later time [when the electron temperature reaches a maximum, see Fig.~\ref{Fig1}(d)], and remains dominant during a comparatively longer time.

\subsection{Plasma Dynamics} When electrons are ejected in large numbers from the surface, they form an electron plasma characterized by a volume density $\rho_{\rm e}(\rb,t)$ that evolves rapidly due to the electromagnetic interaction among the different electrons and the effect of screening by the sample. The latter is driven by the accumulation of positive image charges along the sample surface, with a hole surface density $\sigma_{\rm s}(\rb_s,t)$ depending on the distribution of electrons outside the metal. After an initial fast expansion of the plasma, the attractive interaction between the negatively charged electron cloud and the positively charged surface produces a deceleration in the outgoing motion of the ejected electrons and, eventually, partial reabsorption of plasma electrons. Incidentally, metal screening has a characteristic time \cite{paper004} $\lesssim1$~fs (over which the plasma changes negligibly), and the penetration of the surface screening charge is of the order of the Thomas-Fermi screening length $\lesssim1$~nm; consequently, we model screening in the perfect-metal approximation (see Appendix~\ref{sec:SurfaceScr}).

To realistically describe the spatiotemporal plasma dynamics, we develop a theoretical approach capable of simulating the emission and subsequent evolution of the dense electron plasma cloud, as we describe in Appendix.~\ref{sec:plasmadynamics}. Using this procedure, the simulated spatiotemporal evolution of the plasma density $\rho_{\rm e}(\rb,t)$ and the associated surface hole density $\sigma_{\rm s}(\rb_s,t)$ are both represented in Fig.~\ref{Fig2}(c,e,g,i) as snapshots for selected time delays $\tau$ relative to the time $t_0=0.25~{\rm ps}$ at which the incident pump laser intensity is maximum. 

Right after pumping ($\tau \approx 0~{\rm ps}$), the emission is dominated by the 3PPE channel, and therefore, the plasma accumulates heavily near the sample surface. This occurs because electrons emitted via this process carry a comparatively small kinetic energy $K \approx 3 \hbar \omega-\Phi \sim 10\text{'s}~{\rm meV}$ (see Appendix~\ref{sec:photoemission}), and consequently, they remain relatively close to the surface. As a result, the vast majority of them are rapidly reabsorbed due to image attraction by the metal surface. This behavior is revealed by the sharp peak observed in the green curve of Fig.~\ref{Fig2}(b), representing the number of electrons that remain in free space (i.e., those that have not been reabsorbed) as a function of time, which we compare to the total number of emitted electrons (purple curve). After a few $100\text{'s}~{\rm fs}$, most of the initially emitted electrons have already been reabsorbed. 

Conversely, after this short period, thermionic emission takes over as the dominant emission mechanism, and on average, the so-emitted electrons have a much higher kinetic energy $K \sim 1~{\rm eV}$, producing a noticeable expansion of the plasma cloud up to a few microns from the surface, as revealed by the density maps in Fig.~\ref{Fig2}(c,e,g,i). This evolution is accompanied by fast electron scattering along the wedge side [cf. Figs.~\ref{Fig2}(e) and \ref{Fig2}(g)] due to the transverse asymmetry of the charge distribution, which is strongly concentrated near the tip apex. For this reason, the newly emitted electrons feel a weakened charge barrier when compared to smoother geometries \cite{MDG23}, thus resulting in a faster expansion of the plume. Nevertheless, the intense attractive force of the metal image charges, represented by green lines in Fig.~\ref{Fig2}(c,e,g,i), progressively produces a deceleration and subsequent reversion of the cloud expansion, so that most electrons are eventually reabsorbed. After $10~{\rm ps}$, only $\sim 10~\%$ of the emitted electrons remain in free space, most of which can escape. This amounts to a relatively high portion of escaping electrons, a fact that we explain explained by the aforementioned small charge barrier effect of this particular geometry.

In the calculations presented in this work, we assume the metal structure to be electrically isolated, such that the system maintains charge neutrality (i.e., the number of electrons in the plasma is fully compensated by the number of holes distributed on the metal surface). For grounded samples, additional charges should partially refill the holes, therefore reducing electron reabsorption and affecting the plasma dynamics. We expect this effect to be small for smooth surfaces characterized by large curvature radii, which should produce a screening that is locally approaching the limit of a planar metal surface (i.e., simultaneously meeting the conditions of charge neutrality and a vanishing surface potential).

\section{Results and Discussion}

The charged electron cloud generates an intense electric field in its interior and vicinity. We show in Fig.~\ref{Fig2}(d,f,h,j) the electric field $\mathcal{E}(\rb,t)$ generated by the electron and surface-charge distributions plotted in Fig.~\ref{Fig2}(c,e,g,i), respectively. These maps display a strong concentration of the field amplitude in a region extending up to $\sim 5-10~{\rm \mu m}$ around the tip surface. The field reaches a maximum value right after pumping (i.e., when a large electron pileup is found close to the wedge surface) and slowly dies out as the electrons scatter mainly due to interactions with other electrons in the plasma. Similar dynamics can be observed in a wedge with smaller tip radius (see Fig.~\ref{FigS2}), in which the electrons are emitted from a smaller surface area, thus producing a more concentrated spatial field distribution.

\subsection{Frequency Decomposition of the Generated Fields} 
\begin{figure*}
\centering
\includegraphics[width=0.9\textwidth]{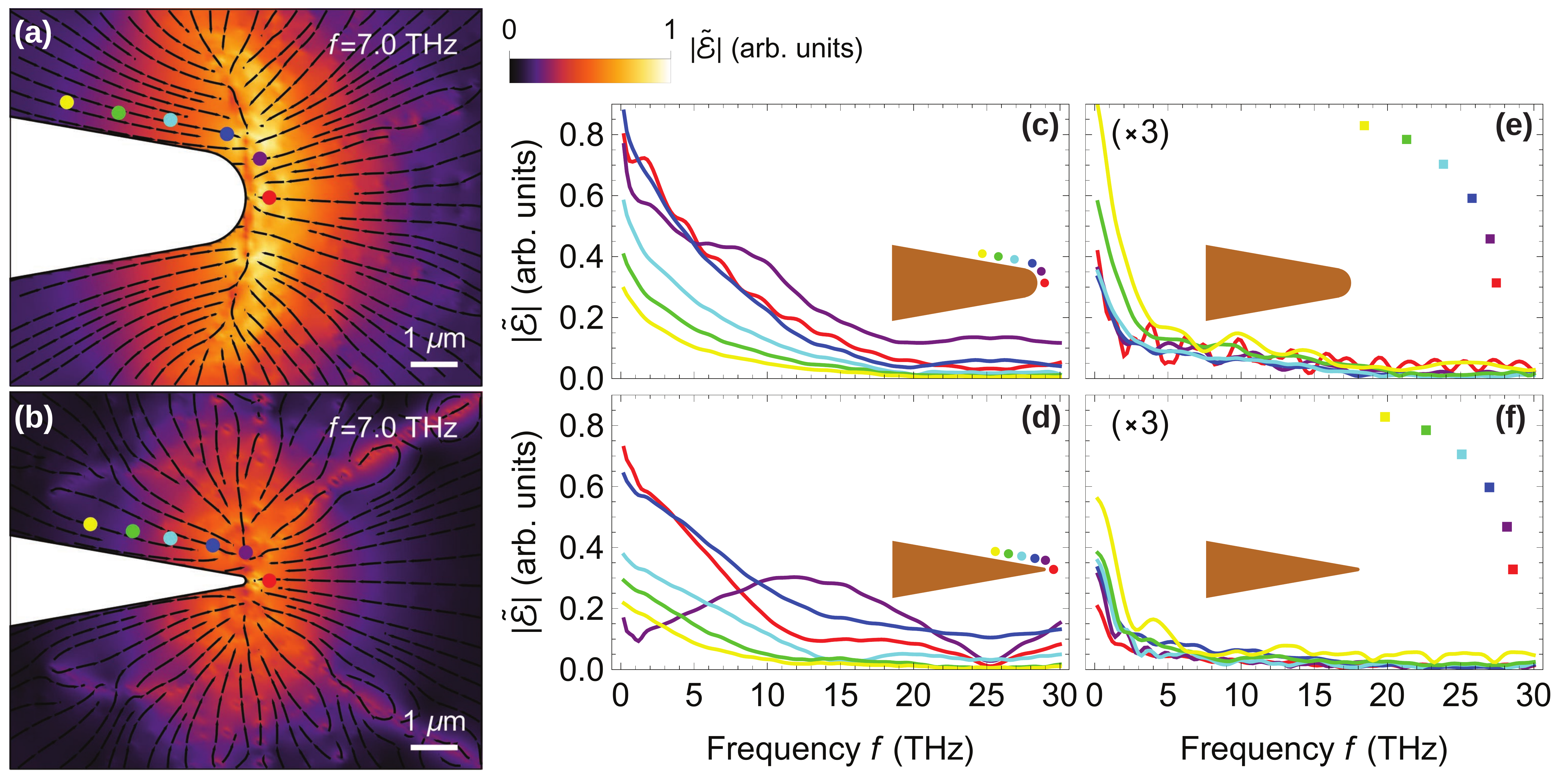}
\caption{{\bf THz electric field in the frequency domain.} {\bf (a,b)} Spatial map of the $f=7.0~\mathrm{THz}$ frequency component of the Fourier-transformed electric field $\tilde{\mathcal{E}}$ near (a) the same copper wedge as in Fig. \ref{Fig2}, with a tip radius $r=0.1~{\rm \mu m}$, and (b) a similar wedge but with $r=100~\mathrm{nm}$ and all other parameters unchanged. The color map represents $|\tilde{\mathcal{E}}|$ and we superimpose field lines (black) tangent to the vector ${\Ree}\{\tilde{\mathcal{E}}\}$. {\bf (c,d)} Spectral decomposition of the electric field around the wedges in (a,b), respectively, at the specific spatial positions marked by color-coordinated dots in the insets [and also in (a,b)], situated at a distance of $500~{\rm nm}$ from the metal surface. {\bf (e,f)} Same as (c,d), but for positions marked by the square dots in the insets, placed at a distance of $10~{\rm \mu m}$ from the metal surface. We set the pump fluence to $F_{\rm pump}=200~\mathrm{mJ/cm^2}$ in all panels.}
\label{Fig3}
\end{figure*}

We perform a spectral analysis of the field produced by the laser-induced plasma by Fourier transforming the time-dependent field [shown in Fig.~\ref{Fig2}(c,e,g,i) for selected instants]. Representative examples of the resulting frequency-domain field $\tilde{\mathcal{E}}(\omega)$ are shown in Figs.~\ref{Fig3}(a) and \ref{Fig3}(b) for wedges with a tip radius of $1$ and $0.1~{\rm \mu m}$, respectively, and for the specific frequency $f=\omega/2\pi=7~{\rm THz}$. In Fig.~\ref{Fig3}(c,d), we represent the full spectral decomposition of the field at selected positions lying at a distance of $0.5~{\rm \mu m}$ from the metal surface, as indicated by color-coordinated dots in Fig.~\ref{Fig3}(a,b), respectively. Upon comparison of the field maps in Fig.~\ref{Fig3}(a,b), we conclude that a smaller tip radius produces a stronger spatial concentration of the near field in the vicinity of the sample, which we explain as the result of electron emission arising from a smaller surface area. Nevertheless, in both cases the field intensity decreases rapidly with distance to the tip, as illustrated in Fig.~\ref{Fig3}(e,f) by plotting the spectral decomposition of the THz far field at several positions separated by a distance of $10~{\rm \mu m}$ from the sample surface. In both wedges, the electric field is dominated by low ${\rm THz}$ frequency components with a mean frequency $\left\langle f \right\rangle \sim 7~{\rm THz}$, although the detailed spectral profile depends on both position and surface geometry. The latter together with the illumination conditions offer potentially useful means to control the generated THz field.

\subsection{Electron Probing of THz Fields} The THz nature of the field observed in Fig.~\ref{Fig3}(b,d) is commensurate with the time scale over which the electron cloud evolves: at a fixed spatial position, there is an initial fast variation in charge density (over the first $\sim 2~{\rm ps}$), responsible for the high-frequency components, followed by a longer period ($\sim 10~{\rm ps}$) in which most electrons have been reabsorbed and the density displays a slow evolution [see Fig.~\ref{Fig2}(b)], giving rise to the sharp increase in low-frequency contributions observed in Fig.~\ref{Fig3}(b,d). This analysis suggests that fast electron pulses traversing the plasma with a controlled delay time relative to the laser pump can serve as excellent probes of the temporal, spectral, and spatial characteristics of the generated THz field.

\begin{figure*}
\centering
\includegraphics[width=0.70\textwidth]{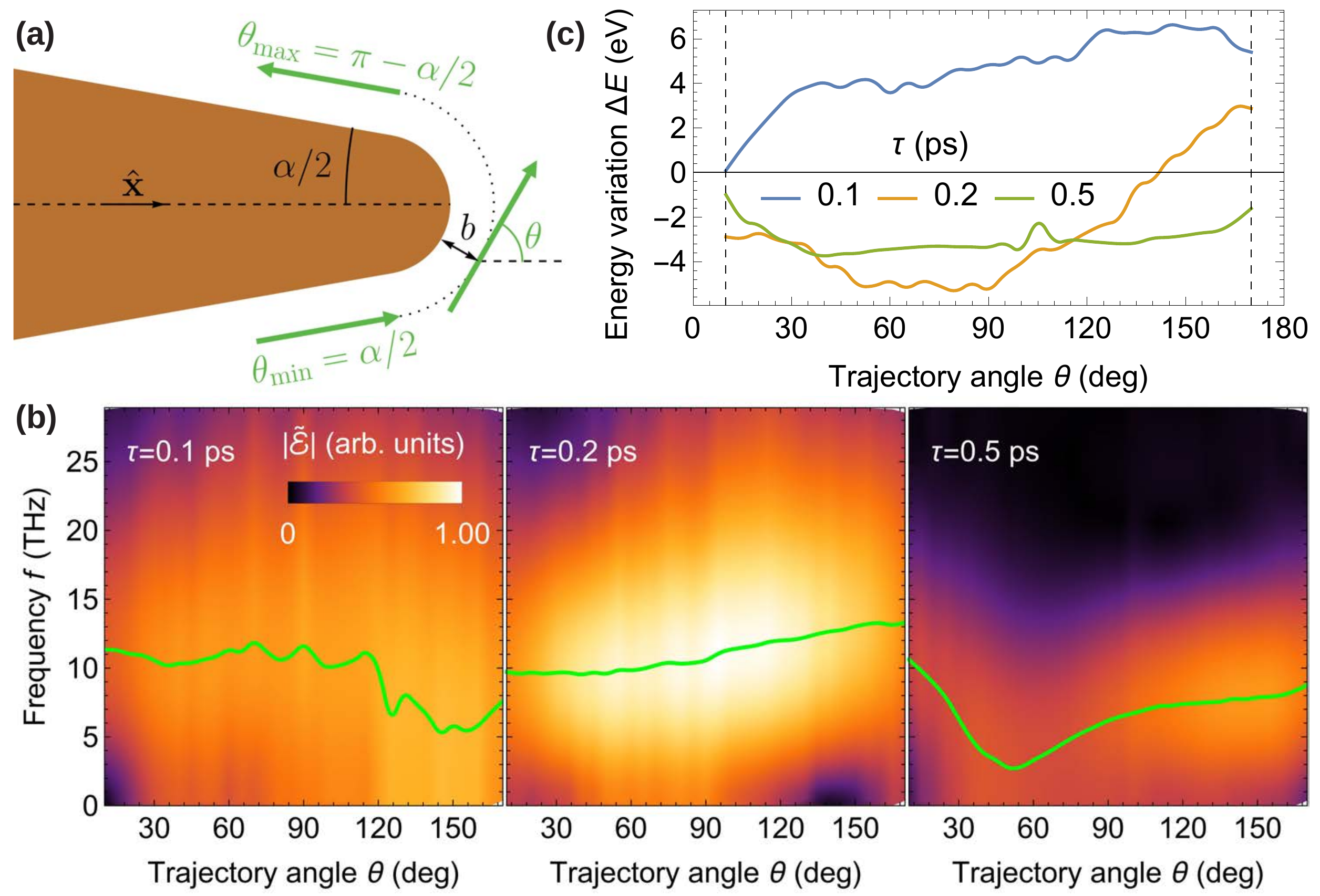}
\caption{{\bf Probing THz fields with e-beam pulses.} {\bf (a)} Scheme showing different electron trajectories (green arrows) passing at a distance $b$ from the surface of the copper wedge considered in Fig.~\ref{Fig2} and forming an angle $\theta$ with the $x$ axis in the $\alpha/2\le\theta\le\pi-\alpha/2$ range, where $\alpha=20^{\circ}$. We introduce a delay $\tau$ between the swift electron and laser pulse. {\bf (b)} Frequency decomposition of the electric field $\mathcal{E}$ along the electron trajectory as a function of frequency $f$ and angle $\theta$ for three different values of $\tau$ (see labels). Green curves represent the peak frequency as a function of $\theta$. {\bf (c)} Net post-interaction electron energy change $\Delta E$ as a function of the trajectory angle $\theta$. The vertical dashed lines indicate the $\theta$ limits $\alpha/2$ and $\pi-\alpha/2$. We take a pump fluence $F_{\rm pump}=200~\mathrm{mJ/cm^2}$, an impact parameter $b=1~\mathrm{\mu m}$, an electron velocity $v_{\rm e}\approx 0.7c$, and an electron pulse duration $\Delta t_{\rm e}=600~\mathrm{fs}$.}
\label{Fig4}
\end{figure*}

To illustrate this idea, we extend our theoretical formalism to incorporate the interaction with a fast probing electron, producing excellent results in comparison with experiments, as shown in recent publications for a different geometry \cite{MDG23,YDG23}. For the wedge structure investigated in this work, we consider an electron passing at a distance $b$ from the surface, with a velocity vector $\vb_{\rm e}$ making an angle $\theta$ relative to the positive $x$ direction, as indicated by the green arrows in Fig. \ref{Fig4}(a) for different values of $\theta$ in the $\alpha/2\le\theta\le\pi-\alpha/2$ range, such that only aloof electron trajectories are considered. We assume the electron wavepacket to be well focused in the transverse e-beam direction and spanning a full-with-half-maximum (FWHM) temporal duration $\Delta t_{\rm e}$. Furthermore, we neglect any changes produced by the interaction on the electron velocity $\vb_{\rm e}$ (nonrecoil approximation, see Appendix~\ref{sec:EnergyVar}).

In Fig. \ref{Fig4}(b), we show the frequency-domain electric field $\tilde{\mathcal{E}}$ amplitude as a function of frequency $f$ and electron trajectory angle $\theta$ for three selected values of the electron delay $\tau$ relative to the pump laser pulse. We set the electron velocity to $v_{\rm e} \approx0.7\,c$ and the impact parameter to $b=1~{\rm \mu m}$. Analogous results for a wedge with smaller tip radius are presented in Fig.~\ref{FigS3}. The maps in both figures confirm that the spectral landscape of the field experienced by the electron depends strongly on both the trajectory angle $\theta$ and the delay $\tau$. In particular, the dominant frequency (i.e., that for which the field intensity is maximum for a given angle $\theta$, represented by the green curves in each plot) can be varied within the $\sim 3-10~{\rm THz}$ range by adjusting these parameters. Likewise, the electron velocity $v_{\rm e}$ and the impact parameter $b$ are additional trajectory parameters that can be varied to explore the frequency landscape, which is also strongly dependent on sample geometry and illumination conditions. As previously reported \cite{MDG23,YDG23} and discussed in Appendix~\ref{sec:EnergyVar}, these electric fields produce a net energy variation $\Delta E$ of the probe electron, which we plot in Fig.~\ref{Fig4}(c) as a function of the trajectory angle $\theta$ for the same values of the delay $\tau$ as in panel (b) (see also Fig.~\ref{FigS3}). The dependence of $\Delta E$ on these parameters is complex because it is mediated by the plasma dynamics during the interaction time. For example, a net gain or loss is observed depending on the electron trajectory. State-of-the-art electron spectrometers can currently resolve energy changes down to $<10$~meV \cite{KDH19}, rendering this approach highly sensitive to minuscule details in the plasma dynamics along the probe electron trajectory.

\section{Conclusions} 

The electron clouds that arise upon irradiation of metallic surfaces with intense laser pulses act as sources of intense THz electromagnetic fields localized over micrometer-sized regions. We have shown that both the spatial extension and the spectral composition of these fields are extremely sensitive to the surface geometry and the characteristics of the pumping. The surface morphology and the illumination conditions are thus elements that can be engineered to control the generated THz fields. These sources could find application in sensing molecular vibrations at similarly low frequencies, with a spatial resolution well below the field wavelength gained by appropriately shaping the metal surface. The localized nature of the generated field is appealing to eliminate spurious signals coming from far regions away from the sample of interest. In addition, the produced THz radiation could be detected without any contamination from the incident laser field, which lies within a completely different spectral region.

The studied process involves the presence of metal holes due to electron ejection. Such holes are redistributed along the surface,  acting as screening charges that strongly affect the plasma dynamics. In this work, we have assumed electrically isolated structures in which charge neutrality leads to a number of holes exactly compensating for the number of plasma electrons. An interesting scenario could be encountered when considering subsequent laser pulses (i.e., impinging on a previously ionized sample), for which we would expect different plasma dynamics under the influence of the net electrostatic potential landscape produced by previous pulses, and eventually, a stationary regime should be reached in which no electrons escape from the structure. A different behavior is also anticipated for grounded samples, in which additional electrons can refill the holes as electrons are ejected away from the surface region. Deviations in the performances of grounded and isolated structures are expected to be stronger in the presence of sharp surface profiles like those considered in this work. We envision the use of an externally controlled degree of insulation (e.g., through a variable resistor) to switch between these two scenarios, thus providing additional means of active control over the generated THz radiation.

As a way to characterize plasma dynamics in this context, we have shown that a passing electron beam pulse with a controlled trajectory can selectively probe specific frequency components, thus offering a unique way to map the spatiotemporal evolution of laser-pulse-induced microplasmas. Probing the ultrafast out-of-equilibrium dynamics of charged-carrier clouds is a challenging problem, whose solution bears interest from both fundamental and applied perspectives. By using the present theory, we have explained recent experiments of spatiotemporal plasma mapping in the context of ultrafast electron microscopy \cite{MDG23}, which have served as a testbed to elucidate the ingredients that play a relevant role in such a complex process, involving different scales of time (from sub-femtosecond metal screening to picosecond plasma evolution), length (from a few nanometers in electron emission and surface charge dynamics to microns in plasma plume dynamics), and energy (from a few electronvolts needed to eject electrons from the metal surface to 100's keV probe electron energies). The effects produced on the probing electron suggest the possibility of designing a disruptive type of micron-sized electron optics component, whereby the wave function associated with free electrons is manipulated by subjecting them to a sizeable and widely controllable interaction with plasma plumes.

\section*{APPENDIX}
\appendix

\section{Two-Temperature Model}
\label{sec:TTM}

We describe the temperature dynamics in a metallic sample irradiated by ultrafast laser pulses through the two-temperature model (TTM), in which the electron and lattice temperatures within the material ($\Te$ and $\Tl$, respectively) are taken as independent variables. For the pulse fluences here considered, the variation in lattice temperature can be neglected ($\Tl\approx T_0$, where $T_0$ is the ambient temperature) and the electron temperature thus obeys the differential equation \cite{paper313}
\begin{align}
    &\ce \frac{d\Te}{dt} = p^{\rm abs} + \nabla\cdot\left(\kappa_{\rm e} \nabla \Te\right) - G(\Te-T_0), \label{eq:TTM}
\end{align}
where $\ce$ is the electron heat capacity, $\kappa_e$ is the electron thermal conductivity, $p^{\rm abs}$ is the power density absorbed from the laser, and $G$ describes electron-phonon coupling.

We calculate the electronic heat capacity of the metal $\ce=\partial Q_{\rm e}(\Te)/\partial \Te$ from the derivative of the temperature-dependent electronic heat density,
\begin{align}
    Q_{\rm e}(\Te)&=\int_{-\infty}^{\infty} dE\, E\,  \rho(E) \left[f_{\mu,\Te}(E)-\Theta(\EF-E)\right],
    \label{eq:I:Qe}
\end{align}
where $\rho(E)$ is the density of states (DOS), $\Theta$ is the step function, $f_{\mu,\Te}(E)=\{\exp[(E-\mu)/(\kB \Te)]+1\}^{-1}$ is the Fermi-Dirac distribution, and $\mu$ is the chemical potential. The latter depends on temperature as determined by the condition
\begin{align}
N_{\rm e}=\int_{-\infty}^{\infty} dE\, \rho(E) f_{\mu,\Te}(E)=\int_{-\infty}^{\EF} dE\, \rho(E),
\label{eq:I:Ne}
\end{align}
expressing the conservation of the number of electrons in the system.

In Eq.~(\ref{eq:TTM}), the absorbed power density at position $\rb$ and time $t$ is given by
\begin{align}
    p^{\rm abs}(\rb,t)=\frac{\omega}{2\pi} \left|\mathcal{E}(\rb,t)\right|^2 \Imm\{\epsilon(\omega)\},
    \label{eq:I:Pabs}
\end{align}
where $\omega$ is the pump frequency, $\epsilon(\omega)$ is the metal permittivity, and $\mathcal{E}(\rb,t)$ is the optical electric field (including scattering by the metal structure). We write the latter as $\mathcal{E}(\rb,t)=\mathcal{E}_{\rm pump} \eta(\rb)\ee^{-(t-t_0)^2/2\overline{\Delta} t^2}$, where $\mathcal{E}_{\rm pump}=\sqrt{2 F_{\rm pump}/(c \overline{\Delta} t)}\pi^{1/4}$ is the pump field amplitude expressed in terms of the fluence $F_{\rm pump}$, $t_0$ marks the time of maximum pulse intensity, we define $\overline{\Delta} t=\Delta t/\sqrt{4\log(2)}$ with $\Delta t$ standing for the FWHM of the intensity, and $\eta(\rb)=\mathcal{E}(\rb,t_0)/\mathcal{E}_{\rm pump}$ is the local field enhancement that we calculate using the boundary-element method\cite{paper040} (BEM).

By assuming that the material surface has a smooth profile characterized by a local curvature radius that is large compared with the light wavelength $\lambda=2\pi c/\omega$, we solve the TTM locally as a 1D model in which any lateral heat diffusion (i.e., along directions parallel to the surface) is neglected and only diffusion along the local direction perpendicular to the surface is considered. This approximation largely simplifies the problem, so that the evolution of the temperature $\Te(\zeta,t)$ (with $\zeta$ standing for the distance from the metal surface towards its interior) can be readily determined from Eq.~\eqref{eq:TTM} using a standard partial differential equation solver.

In this work, we apply this procedure to copper structures, using tabulated data for the DOS of this material \cite{DOSCu} and a Fermi energy $\EF \approx 9.5~\mathrm{eV}$ \cite{DOSCu} corresponding to the chemical potential at $\Te=0$. In addition, we adopt experimental values for the thermal conductivity \cite{keCu} $\kappa_{\rm e}$ and the electron-phonon coupling coefficient\cite{LZC08} $G$. In the present simulations, we set the light wavelength to $\lambda\approx 800~{\rm nm}$, for which the copper permittivity is $\epsilon\approx -25.07+2.54\ii$, and consider pulses with a duration $\Delta t=60~\mathrm{fs}$. Using these parameters to feed the BEM and the TTM, we find the temperature dynamics illustrated in Figs.~\ref{Fig1}(d) and \ref{FigS1}.

\section{Photothermal Electron Emission Mechanisms}
\label{sec:photoemission}

\begin{figure}
\includegraphics[width=0.5\textwidth]{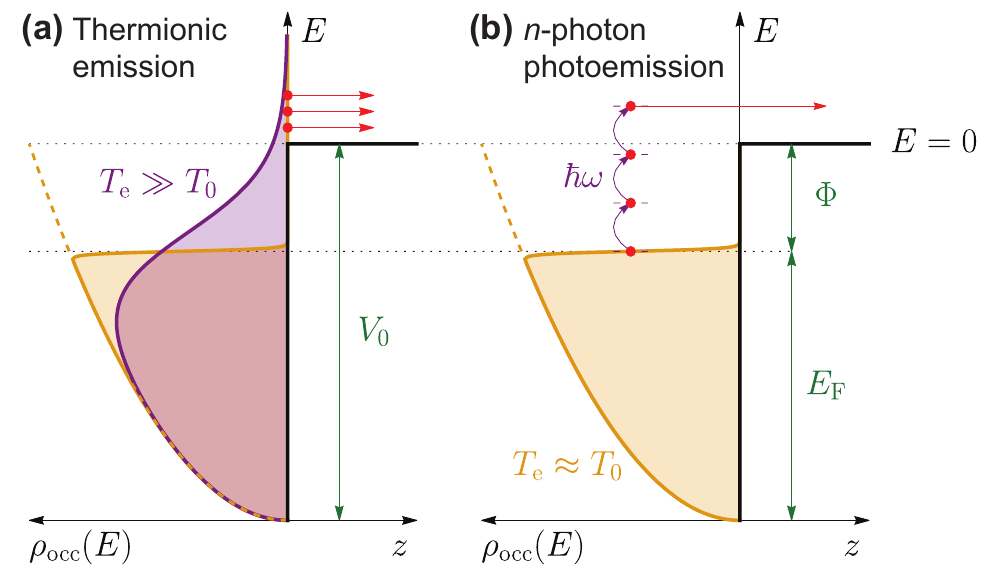}
\caption{{\bf Dominant electron emission mechanisms.} Schematic representation of {\bf (a)} thermionic emission and {\bf (b)} $n$-photon photoemission. In both panels, the dashed orange curve represents the density of states, the filled solid orange curve shows the density of occupied states $\rho_{\rm occ}$ for electrons at room temperature $\Te\approx T_0$, and $z$ stands from the vacuum distance away from the metal surface. The barrier height $V_0$, the Fermi level $\EF$, and the work function $\Phi$ are the same in (a,b). The energy origin $E=0$ is chosen at the barrier top. Electron emission (red arrows) is assisted by an elevated electron temperature $\Te$ in (a), where the filled purple curve shows $\rho_{\rm occ}$ for $\Te\gg T_0$, and by absorption of $n$ photons of energy $\hbar\omega$ in (b) ($n=3$ in the scheme).}
\label{Fig5}
\end{figure}

Under the illumination conditions considered in this work, we assume that electron emission from the metal surface is dominated by two different mechanisms: (1) direct thermionic emission and (2) $n$-photon photoemission. The first of these mechanisms takes place while the metal surface remains hot (for $\gtrsim 1$~ps), such that the elevated electron temperature promotes electrons from lower- to higher-energy states according to the Fermi-Dirac distribution, thus dramatically increasing the electron population for energies above the potential barrier and resulting in electron escape, as depicted in Fig.~\ref{Fig5}(a). In contrast, $n$-photon photoemission occurs only during the pumping time ($<100$ fs) and is driven by the absorption of $n$ photons by one electron, providing it with enough energy to overcome the potential barrier, as depicted in Fig.~\ref{Fig5}(b). We describe each of these two mechanisms below, as well as alternative emission processes.

{\bf Thermionic Emission.} We study this process for an infinite planar surface normal to the $z$ direction, under the approximation of a smooth surface profile (see above). Assuming that conduction electrons inside the metal are confined to a potential well of width $L$ along $z$, the probability per unit area $\mathcal{P}_{\rm th}$ of ejecting an electron across the barrier can be written as
\begin{align}
    \mathcal{P}_{\rm th}(\Te)=\frac{2}{A} \sum_{\kparb k'_z} \mathcal{T}(E_z) f_{\rm FD}(E+V_0,\mu,\Te) (1/\tau) \Theta(E),
    \label{eq:I:PE:P1}
\end{align}
where $A$ is the surface area, the factor of $2$ accounts for spin degeneracy, $\kb=\kparb+k'_z \hat{\bf z}$ is the electron wave vector, $E=\hbar^2 k^2/2m^*-V_0$ is the electron energy, $E_z=E \cos^2\theta$ is the electron energy along $z$, $\theta$ is the emission angle with respect to the $z$ direction,
\begin{align}
    \mathcal{T}(E_z)=\frac{4 \sqrt{E_z} \sqrt{E_z+V_0}}{(\sqrt{E_z}+\sqrt{E_z+V_0})^2} \label{eq:I:PE:TEz}
\end{align}
is the transmittance across the surface energy barrier of height $V_0$, $\mu$ is the chemical potential, $\tau=2L/v$ is the average time interval separating two consecutive electron collisions against the potential barrier, $v=\hbar k_z'/m^*$ is $z$ component of the electron velocity, and $m^*$ is the effective electron mass. Transforming the sums in Eq.~\eqref{eq:I:PE:P1} into integrals through the substitutions $\sum_{\kparb} \to A/(2\pi)^2 \int d\kparb$ and $\sum_{k'_z} \to L/2\pi \int dk_z'$, we obtain
\begin{align}
\mathcal{P}_{\rm th}(\Te) = \int dE \int d\theta\, p_{\rm th}(E,\theta,\Te) \Theta(E),
\label{eq:I:PE:Pe1:final}
\end{align}
where
\begin{align}
    p_{\rm th}(E,\theta,\Te)=\frac{m^*}{2\pi^2\hbar^3} E \cos\theta f_{\rm FD}(E+V_0,\mu,\Te) \mathcal{T}(E \cos^2 \theta)
    \label{eq:I:PE:Pe1:int}
\end{align}
represents the probability of thermionic emission of an electron of energy $E$ along an angle $\theta$ when the electron surface temperature is $\Te$. From here, it follows that electrons are primarily emitted around the surface normal (average emission angle $\theta_{\rm av}=0$) with average energy
\begin{align}
    E_{\rm av}=\frac{\int_{0}^{\infty} dE \int_{-\pi/2}^{\pi/2} d\theta\, E\, p_{\rm th}(E,\theta,\Te)}{\int_{0}^{\infty} dE \int_{-\pi/2}^{\pi/2} d\theta\, p_{\rm th}(E,\theta,\Te)}
    \label{eq:I:th:Eav}
\end{align}
corresponding to an emission velocity $v_{\rm av}=\sqrt{2E_{\rm av}/m_{\rm e}}$.

{\bf $n$-Photon Photoemission.} Under the illumination conditions considered in this work (photon energy $\hbar\omega\approx1.55$~eV, copper work function\cite{K06} $\Phi\approx4.65$~eV), we have $4>\Phi/\hbar\omega \gtrsim 3$, so that this emission channel is dominated by $n=3$ processes. The corresponding photoemission rate is calculated using the well-known Fowler-Dubridge model, according to which the emission probability is given by \cite{FBG09}
\begin{align}
    \mathcal{P}_{3{\rm PPE}}(\Te)=a_3 \mathcal{A} \left(\frac{e}{\hbar \omega}\right)^3 I_{\rm abs}^3 T_{\rm e}^2 F\left(\frac{3\hbar\omega-\Phi_{\Te}}{\kB \Te}\right),
    \label{eq:I:PE:Pe2:final}
\end{align}
where $\mathcal{A}=120\ \mathrm{A/cm^2K^2}$ is the Richardson constant, $I_{\rm abs}$ is the absorbed power density, $\Phi_{\Te}=\Phi+\EF-\mu$ is the temperature-corrected work function, $F(x)=\int_0^{\infty}dy \log[1+\exp(x-y)]$ is the Fowler function \cite{FBG09}, and $a_3\sim 5\times 10^{-36}\ {\rm cm^6/A^{3}}$ \cite{MDG23} represents the likelihood of the emission. Finally, the average energy distribution of the photoemitted electrons can be calculated as
\begin{align}
    E_{\rm av}=\frac{\int_{0}^{\infty}dE\, E\, \rho_{\rm occ}(E-n\hbar\omega)}{\int_{0}^{\infty}dE\, \rho_{\rm occ}(E-n\hbar\omega)},
    \label{eq:I:nppe:Eav}
\end{align}
while the average emission angle is again $\theta_{\rm av}=0$ due to symmetry.

{\bf Alternative Emission Mechanisms.} Under strong field illumination, conduction electrons could escape from the metal via tunneling into the vacuum due to the periodic lowering of the potential barrier by the incident laser electric field. According to the Keldysh criterion, this mechanism is negligible compared to $n$-photon photoemission if $\gamma\ll1$, where 
\begin{align}
    \gamma=\sqrt{\frac{\Phi}{2 U_{\rm p}}}=\frac{\omega \sqrt{2m_{\rm e}\Phi}}{e |\mathcal{E}|}
    \label{eq:I:Keldysh}
\end{align}
is the dimensionless Keldysh parameter\cite{DPV20,K1965}, defined in terms of the metal work function $\Phi$ and the ponderomotive energy $U_{\rm p}=e^2 |\mathcal{E}|^2/4m_{\rm e}\omega^2$, which is in turn expressed in terms of the light frequency $\omega$ and the electric field amplitude $|\mathcal{E}|$. For the copper samples considered in this work, under illumination by a $800~{\rm nm}$ laser with a peak of intensity of $\sim 300-1500~{\rm GW/cm^2}$, we have $\gamma\sim 10-20 \gg 1$, and consequently, we neglect tunneling emission. Another possible emission mechanism is the escape of nonthermal high-energy electrons during a short period right after pumping when the system is strongly out of equilibrium. However, we expect this contribution to only amount to a small correction in the total emission, and thus, we neglect it as well.

\section{Plasma Dynamics}
\label{sec:plasmadynamics}

We now describe our numerical implementation to simulate the emission and spatial evolution of the electron plasma, starting with a discretization of the surface through a set of positions $s_j$, and also the time intervals $t_i$ at which electrons have been emitted, where $i$ and $j$ are discretization indices. Electrons within each set of $(i,j)$ indices are evolved independently, taking into consideration the interaction with both surface charges and other sets. We are interested in the evolution of the corresponding densities of emitted electrons $\rho_{ij}(t)$, and further represent the dynamics of each $(i,j)$ set with a time-dependent average velocity $\mathbf{v}_{ij}(t)$. The initial population of every $(i,j)$ set is determined by the emission probabilities in Eq.~\eqref{C6:eq:Pe:final}, which are described in Appendix~\ref{sec:photoemission}. To alleviate the computational burden, we consider all electrons to be ejected normally to the local surface with a velocity $v_{\rm av}=\sqrt{2 E_{\rm av}/\me}$ determined by the corresponding average energy $E_{\rm av}$ (see above), which is, in turn, dependent on $(i,j)$ through the local field amplitude and electron temperature.

To compute the dynamical evolution of each $(i,j)$ set, we need to calculate the force acting on the electrons at any given time $t\geq t_i$:
\begin{align}
    \mathbf{F}_{ij}(t) = \sum_{i'\leq i} \sum_{jj'} \mathbf{f}^{\rm ee}_{ij,i'j'}(t) +  \sum_{\ell} \mathbf{f}^{\rm eh}_{ij,\ell}(t),
\end{align}
where $\mathbf{f}^{\rm ee}_{ij,i'j'}(t)$ refers to the electron-electron (ee) interaction with the rest of the previously emitted $(i',j')$ sets, while $\mathbf{f}^{\rm eh}_{ij,\ell}(t)$ is the contribution of surface charges [electron-hole (eh) interaction; see Appendix~\ref{sec:SurfaceScr}] summed over surface positions $s_{\ell}$. Neglecting magnetic interactions due to the small drift velocity of the emitted electrons, the ee force component is given by
\begin{subequations}
\begin{widetext}
\begin{align}
    \mathbf{f}^{\rm ee}_{ij,i'j'}(t) &=  \int_{-D_y/2}^{D_y/2} dy \int_{-D_y/2}^{D_y/2} dy' \frac{e^2 \rho_{ij}(t_{\rm r}) \rho_{i'j'}(t_{\rm r}) (x_{ij}-x_{i'j'},y-y',z_{ij}-z_{i'j'})}{[(x_{ij}-x_{i'j'})^2+(y-y')^2+(z_{ij}-z_{i'j'}))^2]^{3/2}} \nonumber\\
    &= 2 e^2 \rho_{ij}(t_{\rm r}) \rho_{i'j'}(t_{\rm r}) \frac{\rb_{ij}-\rb_{i'j'}}{|\rb_{ij}-\rb_{i'j'}|^2} \left[ \sqrt{D_y^2+|\rb_{ij}-\rb_{i'j'}|^2}-|\rb_{ij}-\rb_{i'j'}| \right],
    \label{eq:fee}
\end{align}
\end{widetext}
where $\rb_{ij}=(x_{ij},0,z_{ij})$ and we introduce the retarded time of interaction $t_{\rm r}=t-|\rb_{ij}-\rb_{i'j'}|/c$. We find that this retardation correction affects the results because of the large extension of the plume, which is not negligible compared with the wavelength associated with the generated THz field. It should be noted that the $y$ and $y'$ integrals are needed because $\rho_{\rm e}$ (and also the surface charge density, see below) is defined per unit length along that direction and we assume translational invariant in both the geometry and the pump. To connect with experiments, in which the pump laser beam has a finite lateral extension, we have introduced a parameter $D_y$ accounting for an effective length along $y$, within which we approximate the density of plasma electrons (and also surface charges, see below) to be constant. We set $D_y=25~{\rm \mu m}$ in the present calculations. Analogously, the eh contribution reads
\begin{align}
    \mathbf{f}^{\rm eh}_{ij,\ell}(t) = -2 e^2 \rho_{ij}(t_{\rm r}) \sigma_{\ell}(t_{\rm r}) \frac{\rb_{ij}-\rb_{\ell}}{|\rb_{ij}-\rb_{\ell}|^2} \nonumber \\ \times \left[ \sqrt{D_y^2+|\rb_{ij}-\rb_{\ell}|^2}-|\rb_{ij}-\rb_{\ell}| \right],
    \label{eq:feh}
\end{align}
\end{subequations}
where $\sigma_{\ell}$ is the density of holes per unit length along $y$ within the surface interval represented by the point $s_{\ell}$ of coordinates $\rb_{\ell}=(x_{\ell},0,z_{\ell})$, and $t_{\rm r}=t-|\rb_{ij}-\rb_{\ell}|/c$.

The corresponding acceleration that this force exerts on electrons in the $(i,j)$ set is given by $\mathbf{a}_{ij}(t) = \mathbf{F}_{ij}(t) / M_{ij}(t)$, where we assimilate $M_{ij}(t)=\me D_y\,\rho_{ij}(t)$ to the total mass of a uniform charged line with extension $D_y$ along $y$ and mass density $\me \rho_{ij}(t)$. The velocity and position are then updated at each time step $\delta t$ according to Newton's equation as $\vb_{ij}(t+\delta t) = \vb_{ij}(t)+\mathbf{a}_{ij}(t)\delta t$ and  $\rb_{ij}(t+\delta t) = \rb_{ij}(t)+\mathbf{v}_{ij}(t)\delta t$, respectively. Simultaneously evolving all $(i,j)$ sets, we construct the time-dependent electron density $\rho_{\rm e}(\rb,t)=\sum_{ij} \rho_{ij}(t)\delta[\rb-\rb_{ij}(t)]$, from which the surface hole distribution $\sigma_\ell(t)$ is also updated at each time $t$ using the method described in Appendix~\ref{sec:SurfaceScr}.

When electrons move back to the surface, such that $\rb_{ij}(t+\delta t)$ is located inside the metal at time $t+\delta t$ (but outside at time $t$), we introduce the effect of electron reabsorption and partial reflection by considering that a fraction of the arriving electrons is specularly reflected. This is done by inverting the sign of the surface-normal component of the velocity $v_{ij}^{\perp}$ and making $\rho_{ij}(t+\delta t)=\rho_{ij}(t) [1-\mathcal{T}(E_{ij}^{\perp})]$, where $E_{ij}^{\perp}=\me (v_{ij}^{\perp})^2/2$ is the normal electron energy and $\mathcal{T}$ is a transmittance coefficient given by Eq.~\eqref{eq:I:PE:TEz}. Since $1-\mathcal{T}(E_{ij}^{\perp})<1$, this procedure produces a depletion in the number of plasma electrons (i.e., reabsorption).

\section{Surface Screening Charge}
\label{sec:SurfaceScr}

We approximate the metal as a perfect conductor because screening has a characteristic time and length of \cite{paper004} $\lesssim1$~fs and $\lesssim1$~nm, much smaller than the spatiotemporal scales involved in the formation and evolution of the plasma. The screening charge is then obtained by an adaptation of the boundary-element method for perfect conductors \cite{paper194}. Taking a structure that is translationally invariance along $y$, we consider the transverse profile $\rb_s=(x_s,0,z_s)$, parametrized by a variable $s$ that evolves linearly from 0 to 1 as we go around the perimeter length $P$. We now consider a line of charge aligned along $y$, placed at a transverse position $\rb_c=(x_c,0,z_c)$, and having a charge density $q$ per unit length. The presence of the external charge places the metal surface at a potential $V$, which is uniform in the limit of a perfect conductor. Then, the distribution of induced surface charges $e\sigma_s$ (charge per unit of surface area) is determined by the condition
\begin{align}
V=\int_{-\infty}^\infty &dy\;\bigg[\frac{q}{\sqrt{(x_s-x_c)^2+(z_s-z_c)^2+y^2}} \nonumber\\
&+eP\int_0^1 ds'\;\frac{\sigma_{s'}}{\sqrt{(x_s-x_{s'})^2+(z_s-z_{s'})^2+y^2}}\bigg],
\nonumber
\end{align}
which needs to be satisfied at all positions $s$. The $y$ integral of each of the fractions in this expression produces a logarithmic divergence at large distances. However, the overall divergence cancels due to the neutrality of the total external plus induced charges, and thus, upon integration, we find
\begin{align}
&-V=q\,\log\bigg\{\big[(x_s-x_c)^2+(z_s-z_c)^2\big]\big/P^2\bigg\} \nonumber\\
&+eP\int_0^1 ds'\;\sigma_{s'}\,\log\bigg\{\big[(x_s-x_{s'})^2+(z_s-z_{s'})^2\big]\big/P^2\bigg\}.
\nonumber
\end{align}
Incidentally, we normalized the arguments of the logarithms by dividing by $P^2$ to obtain dimensionless quantities, but any normalization factor in these functions cancels because of charge neutrality. We solve this equation by discretizing the transverse surface profile through a set of $N$ equally spaced points corresponding to $s_\ell=(\ell+1/2)/N$, with $\ell=0,\cdots,\,N-1$, leading to the linear equation $M\cdot\sigma=b$, where we define an $N\times N$ matrix of components $M_{\ell,\ell'}=(eP/N)\log\big\{\big[(x_{s_\ell}-x_{s_{\ell'}})^2+(z_{s_\ell}-z_{s_{\ell'}})^2\big]\big/P^2\big\}$, as well as the $N$-vectors $\sigma_\ell=\sigma_{s_\ell}$ and $b_\ell=-q\,\log\big\{\big[(x_{s_\ell}-x_c)^2+(z_{s_\ell}-z_c)^2\big]\big/P^2\big\}$.

For a biased structure, the potential $V$ is taken as a parameter (e.g., $V=0$ for grounded samples) and the screening charge in the presence of such potential is just obtained upon inversion of the aforementioned $N\times N$ linear system of equations.

In this work, we consider instead electrically isolated metal structures, so we need to impose charge neutrality through the equation $eP\int_0^1 ds\;\sigma_{s}=-q$. Then, the potential $V$ is no longer a parameter, but rather a variable determined by the new equation. Consequently, the above matrix and vectors need to be supplemented with additional components $M_{N,\ell}=eP/N$, $M_{\ell,N}=1$, $M_{N,N}=0$, $b_N=-q$, and $\sigma_N=V$, thus defining an enlarged $(N+1)\times(N+1)$ system.

In both scenarios (grounded and isolated structure), we need to invert the corresponding system of linear equations to find the surface hole distribution $\sigma_s$ for a line charge $q$ placed at $\rb_c$. At each time $t$ along the evolution of the plasma, we then calculate the total surface charge as the superposition of those generated by all plasma elements $q=-e\rho_{ij}(t)$ at positions $\rb_c=\rb_{ij}(t)$.

\section{Energy Variation of the Probe Electron}
\label{sec:EnergyVar}

When an electron with energy $E_0$ and velocity $\vb_{\rm e}$ passes by the vicinity of the metal structure, it interacts with the plasma and the sample, thus undergoing a variation in energy by an amount
\begin{align}\label{C6:eq:DeltaE}
    \Delta E= -e \vb_{\rm e}\cdot \int_{-\infty}^{\infty} dt\; \mathcal{E}[\rb_{\rm e}(t),t] \equiv \int_{-\infty}^{\infty} dt\; \Gamma(t),
\end{align}
where $\mathcal{E}[\rb_{\rm e}(t),t]$ is the electric field on the trajectory of the electron $\rb_{\rm e}(t)$, we define $\Gamma(t)$ as the electron energy variation rate, as we adopt the nonrecoil approximation (i.e., $\vb_{\rm e}$ is constant and $|\Delta E|\ll E_0$). The classical energy change represented by Eq.~(\ref{C6:eq:DeltaE}) is a good approximation even when considering electrons as quantum wavepackets, as shown in Refs.~\citenum{MDG23,YDG23}. The electric field is given by $\mathcal{E}(\rb,t) = -{\bf \nabla}\phi(\rb,t)$, where $\phi=\phi_{\rm e}+\phi_{\rm h}$ is the potential generated by the plasma electrons ($\phi_{\rm e}$) and the induced surface charges ($\phi_{\rm h}$). We ignore the effect of the vector potential due to the low drift velocity of the emitted electrons. Correspondingly, the electric field can be separated into the contributions arising from the emitted plasma electrons,
\begin{subequations}\label{C6:eq:Geh}
\begin{align}
    &\mathcal{E}_{\rm e}(\rb,t)
    = \nonumber \\ &-2e D_y \int dx'\int dz' \, \rho_{\rm e}(\rb',t_{\rm r}) \frac{\rb - \rb'}{|\rb - \rb'|^2 \sqrt{D_y^2 + 4|\rb - \rb'|^2}},\label{C6:eq:Ge}
\end{align}
and the associated induced surface charges,
\begin{align}
    &\mathcal{E}_{\rm h}(\rb,t)= \nonumber\\
    & 2e P D_y \int_0^1 ds\, \sigma_{\rm s}(s,t_{\rm r}) \frac{\rb - \rb_s}{|\rb - \rb_s|^2 \sqrt{D_y^2 + 4|\rb - \rb_s|^2}}, \label{C6:eq:Gh}
\end{align}
\end{subequations} 
where $\rb=(x,0,z)$, $\rb'=(x',0,z')$, $\rb_s=(x_s,0,z_0)$ runs over surface positions parameterized by $s$, $P$ is the perimeter of the metal cross section, $\rho_{\rm e}$ and $\sigma_{\rm s}$ are the densities of emitted electrons and surface holes, respectively [see Fig.~\ref{Fig2}(c,e,g,i) in the main text], and $t_{\rm r}$ is the retarded time defined as $t-|\rb - \Rb|/c$ in Eq.~\eqref{C6:eq:Ge} and $t-|\rb - \rb_s|/c$ in Eq.~\eqref{C6:eq:Gh}. The e-beam energy variation rates can equally be separated into the corresponding contributions as $\Gamma_{\rm e/h}(t)=-e \vb_{\rm e}\cdot \mathcal{E}_{\rm e/h}[\rb_{\rm e}(t),t]$. To account for the finite FWHM of the electron wavepacket $\Delta t_{\rm e}$, we correct Eqs.~\eqref{C6:eq:Ge} and \eqref{C6:eq:Gh} by performing a Gaussian convolution with the same FWHM, such that $\Gamma(t)$ is replaced by $\Gamma^{\rm av}(t) = (1/\overline{\Delta} t_{\rm e}\sqrt{\pi})\int_{-\infty}^{\infty} dt' \ee^{-(t-t')^2/\overline{\Delta} t_{\rm e}^2}  \left.\Gamma(t)\right|_{\rb_{\rm e}(t) \to \rb_{\rm e}(t')}$ with $\overline{\Delta} t_{\rm e}=\Delta t_{\rm e}/\sqrt{4\log(2)}$.

In the numerical implementation of the calculation of the e-beam energy variation, the electron plasma distribution at each time $t$ is discretized through a uniform grid of element size $\delta x \times \delta z$, constructed such that all electrons placed inside each grid element are assimilated to a single effective charge, with a linear (along $y$) density given by the sum of those associated with the enclosed electrons. Analogously, the surface is also discretized with elements of equal length $\delta s$ along the transverse surface profile, each of them containing an effective positive image charge. This procedure runs smoothly when evaluating the field (and the induced forces) at large distances by simply placing the effective charges 
at the center of the grid or surface elements. However, extra care needs to be taken at short distances, for which we consider each grid element to be uniformly charged. Equations.~\eqref{C6:eq:Geh} are then corrected by performing the transformations
\begin{subequations}
\begin{align}
    \tilde{\mathcal{E}}_{\rm e}(\rb,t) = \frac{1}{\delta x \delta y} \int_{x-\delta x/2}^{x+\delta x/2} dx' \int_{z-\delta z/2}^{z+\delta z/2} dz'\, \mathcal{E}_{\rm e}(\rb',t)
\end{align}
and
\begin{align}
    \tilde{\mathcal{E}}_{\rm h}(\rb_s,t) = \frac{1}{\delta s} \int_{s-\hat{\mathbf{t}}\delta s/2}^{s+\hat{\mathbf{t}}\delta s/2} ds'\, \mathcal{E}_{\rm h}(\rb_{s'},t),
\end{align}
\end{subequations}
where $\hat{\mathbf{t}}$ is a surface-tangent vector at the $s$-dependent position $\rb_s$. To evaluate these integrals, we can safely neglect the effect of the grid size on the terms $\sqrt{D_y^2 + 4|\rb - \Rb|^2}$ and $\sqrt{D_y^2 + 4|\rb - \rb_s|^2}$ in Eqs. \eqref{C6:eq:Ge} and \eqref{C6:eq:Gh}, respectively, as we have $D_y \gg \sqrt{\delta x^2+\delta y^2}$ and $D_y \gg P \delta s$; therefore, we only need to compute the integrals
\begin{subequations}
\begin{align}
    &\frac{1}{\delta x \delta y} \int_{x-\delta x/2}^{x+\delta x/2} dx' \int_{z-\delta z/2}^{z+\delta z/2} dz'\, \frac{\rb'}{|\rb'|^2}= \nonumber \\ &\frac{2}{\delta x \delta z} 
    \begin{bmatrix}
g(x_+,z_+)-g(x_+,z_-)-g(x_-,z_+)+g(x_-,z_-)\\
g(z_+,x_+)-g(z_+,x_-)-g(z_-,x_+)+g(z_-,x_-)
\end{bmatrix} \label{eq:Corr:e}
\end{align}
and 
\begin{align}
    &\frac{1}{\delta s} \int_{s-\hat{\mathbf{t}}\delta s/2}^{s+\hat{\mathbf{t}}\delta s/2} ds'\, \frac{\rb_{s'}}{|\rb_{s'}|^2} = \nonumber\\ &\frac{2}{\delta s} \left[ \frac{\rb_s-(\rb_s \cdot \hat{\mathbf{t}})\hat{\mathbf{t}}}{\Delta}h_1(\rb_s)+\frac{\hat{\mathbf{t}}}{2}h_2(\rb_s) \right], \label{eq:Corr:h}
\end{align}
\end{subequations}
where we define $x_\pm = x \pm \delta x/2$, $z_\pm = z \pm \delta z/2$, $\Delta=\sqrt{r_s^2-(\rb_s \cdot \hat{\mathbf{t}})^2}$,
\begin{subequations}
\begin{align}
    g(x,z)=x \arctan( z/x ) + (z/2) \log(x^2+z^2),
\end{align}
\begin{align}
    &h_1(\rb_s)=\nonumber\\&\arctan\left( \frac{\rb_s \cdot \hat{\mathbf{t}}+\delta s/2}{\Delta} \right)-\arctan\left( \frac{\rb_s \cdot \hat{\mathbf{t}}-\delta s/2}{\Delta} \right),
\end{align}
and
\begin{align}
    h_2(\rb_s)=\log\left(\frac{|\rb_s \cdot \hat{\mathbf{t}}+\delta s/2|}{|\rb_s \cdot \hat{\mathbf{t}}-\delta s/2|}\right);
\end{align}
\end{subequations}
finally, this correction is applied by using Eqs.~\eqref{eq:Corr:e} and \eqref{eq:Corr:h} to replace the terms $(\rb-\Rb)/|\rb-\Rb|^2$ and $(\rb-\rb_s)/|\rb-\rb_s|^2$ in Eqs. \eqref{C6:eq:Ge} and \eqref{C6:eq:Gh}, respectively.



\begin{acknowledgments} 
This work has been supported in part by the European Research Council (Advanced Grant 789104-eNANO and Starting Grant 851780-NanoEP), the European Commission (Horizon 2020 Grant 964591-SMART-electron), the Spanish MICINN (PID2020-112625GB-I00 and Severo Ochoa CEX2019-000910-S), Google Inc., the Catalan CERCA Program, and Fundaci\'{o}s Cellex and Mir-Puig. 
\end{acknowledgments} 


%

\pagebreak
\onecolumngrid
\section*{SUPPLEMENTARY FIGURES}

\section{Supplementary Figures}
\label{sec:SI}

\begin{figure*}[b]
\begin{centering}
\includegraphics[width=1.00\textwidth]{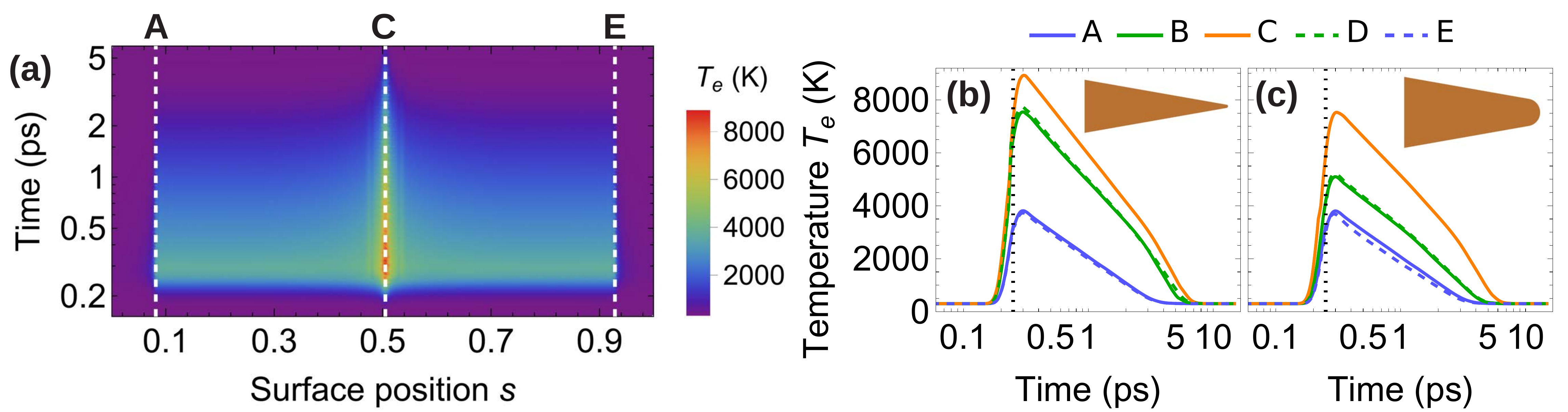}
\par\end{centering}
\caption{\textbf{Sample temperature dynamics.} {\bf (a)} Same as Fig.~\ref{Fig1}(d) in the main text, but for a wedge with tip radius $r=0.1~{\rm \mu m}$ and all remaining parameters unchanged. The vertical dashed lines mark the positions of the blue dots in Fig.~\ref{Fig1}(c) along the surface of the wedge, with points B and D omitted, as they lay very close to point C in this geometry. {\bf (b,c)} Temperature profile as a function of time at the positions A--E in the wedges studied in (a) and Fig.~\ref{Fig1}(d) of the main text, respectively. The dotted black lines in (b,c) represent the time $t_0=0.25~{\rm ps}$ of maximum incident pump laser intensity.}
\label{FigS1}
\end{figure*}

\begin{figure*}
\begin{centering}
\includegraphics[width=0.7\textwidth]{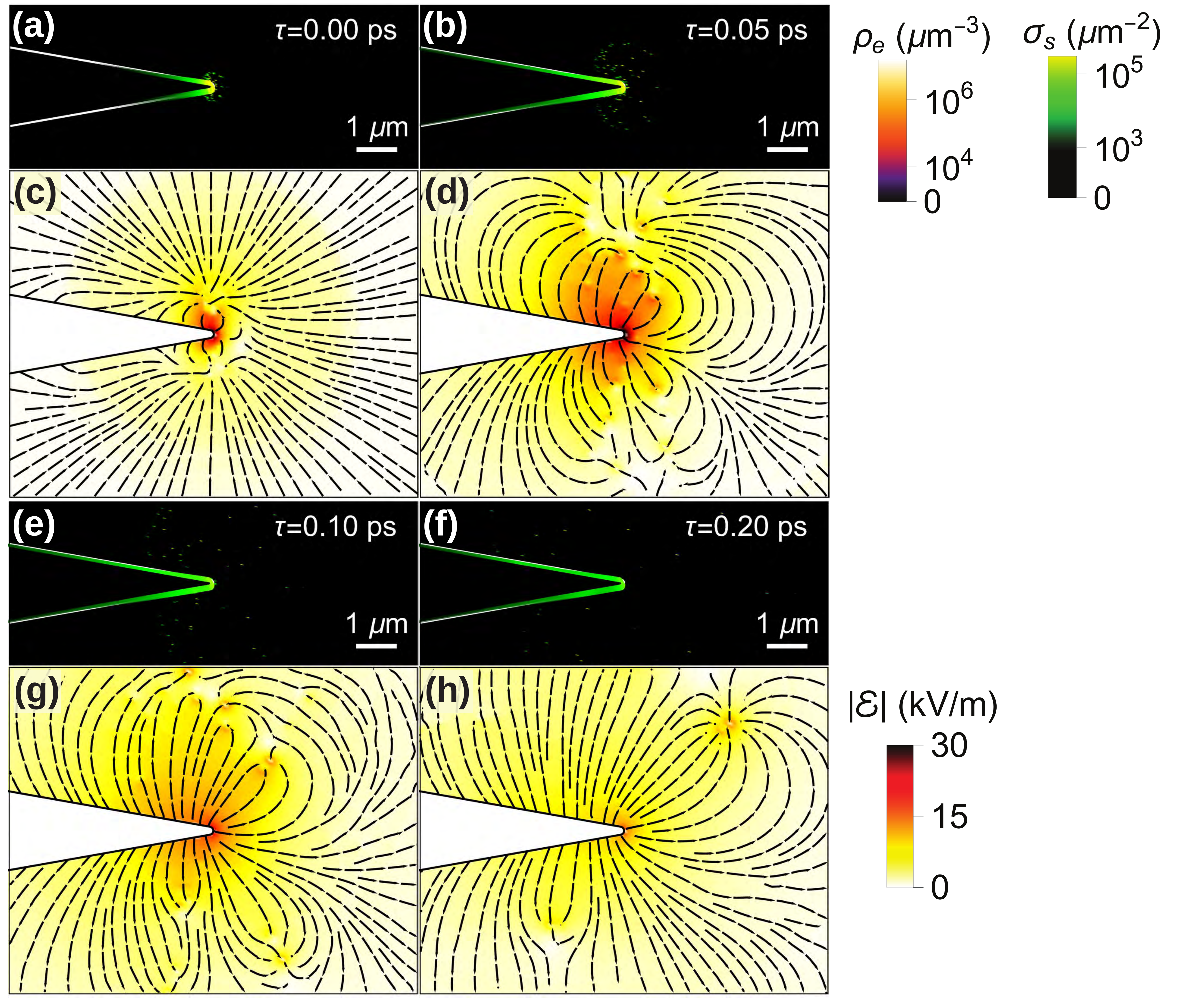}
\par\end{centering}
\caption{{\bf Plasma-induced electric field.} Same as Fig.~\ref{Fig2}(c-j) in the main text, but for a wedge with tip radius $r=0.1~{\rm \mu m}$ and all remaining parameters unchanged.}
\label{FigS2}
\end{figure*}

\begin{figure*}
\begin{centering}
\includegraphics[width=0.8\textwidth]{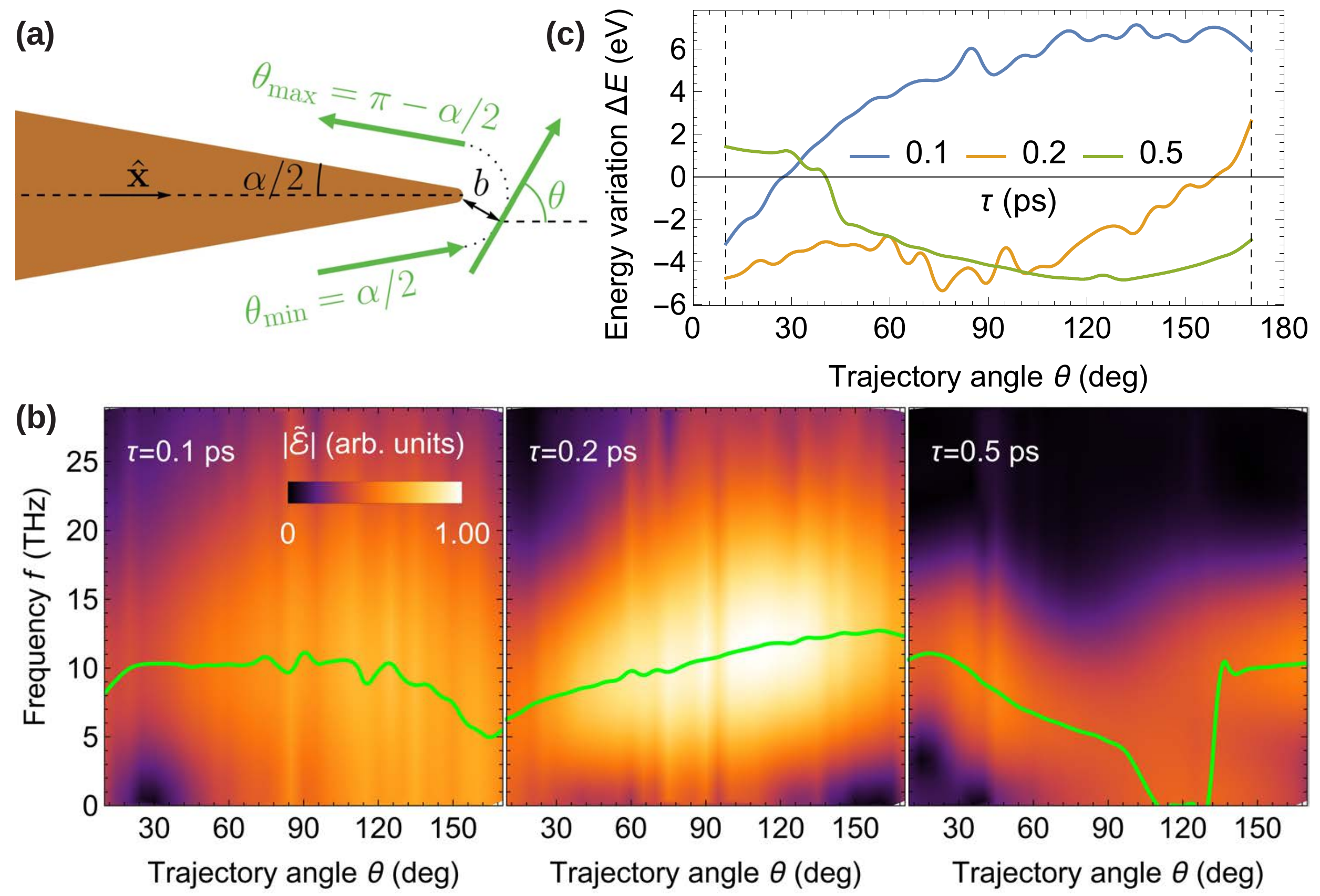}
\par\end{centering}
\caption{{\bf Probing THz fields with e-beam pulses.} Same as Fig.~\ref{Fig4} in the main text, but for a wedge with tip radius $r=0.1~{\rm \mu m}$ and all remaining parameters unchanged.}
\label{FigS3}
\end{figure*}

We present supplementary figures containing additional details of the electron temperature dynamics (Fig.~\ref{FigS1}) and analogous results to Figs.~\ref{Fig2} and \ref{Fig4} (Figs.~\ref{FigS2} and \ref{FigS3}, respectively), but for a wedge with a smaller tip radius.

\end{document}